\documentstyle[twoside,fleqn,espcrc2,epsfig]{article}

\title{Considerations on Neuberger's operator.}
 
\author{L.~Giusti$^{\rm a}$, Ch.~Hoelbling$^{\rm a}$ and C.~Rebbi
\address{Boston University Physics Department\\
         590 Commonwealth Avenue\\
         Boston MA 02215, USA}
\thanks{Presented by C.~Rebbi. This research was supported in part 
under DOE grant DE-FG02-91ER40676.}}

\begin{document}

\begin{abstract}
We discuss new approaches to the numerical implementation of 
Neuberger's operator for lattice fermions and the possible use 
of block spin transformations.
\end{abstract}

\maketitle

Recently major progress has been made in the lattice discretization of
fermionic fields.  Starting from the overlap formulation, Neuberger
has proposed a lattice Dirac operator which avoids the additive
renormalization of fermion masses~\cite{Neuberger1}.  This operator
satisfies an identity, the Ginsparg-Wilson relation~\cite{GW}, by which
one can define exact lattice chiral symmetry~\cite{Luescher1}. Other
lattice operators which satisfy the Ginsparg-Wilson relations have
been obtained as implementations of the ``perfect action''~\cite{HLN}.
In this note we will concentrate on Neuberger's operator and we
will follow his notation.  If one defines the Wilson lattice operator
as
\begin{eqnarray*}
  (D_W\psi)(x)=\frac{\psi(x)}{\kappa}-\sum_{\mu}[(1-\gamma_{\mu})
  U_{\mu}(x) \psi(x+\mu)
\end{eqnarray*}
\begin{equation}
  \qquad \qquad +(1+\gamma_{\mu})U_{\mu}(x-\mu) \psi(x-\mu)]
  \label{eq:1}
\end{equation}
Neuberger's operator is 
\begin{equation}
  D=\frac{1}{2}(1+V)    
  \label{eq:2}
\end{equation}
where the unitary operator $V$ is given by
\begin{equation}
  V=D_W (D_W^{\dag} D_W)^{-1/2}
  \label{eq:3}
\end{equation}
Provided the hopping parameter $\kappa$ in Eq.~\ref{eq:1} is chosen within
a suitable range $(1/8 - 1/4)$, Neuberger's operator defines a single
flavor of massless lattice fermions, symmetric under chiral transformations.

Since in actual calculations the size of the matrix $D_W$ is very
large, for a practical implementation of Neuberger's operator it is
crucial to find computationally efficient ways to calculate its action
as well as the action of its inverse on any given vector.  In some sense,
the domain wall formulation of lattice fermions~\cite{dw1,dw2,dw3}, 
which preceded the introduction of the operator of Eq.~\ref{eq:2}, 
provide a numerical procedure for its implementation.  Other approximate 
methods have been proposed in~\cite{Neuberger2,EHN,Luescher2,Borici}.  
Most of these numerical methods proceed through an approximation to 
the operator
\begin{equation}
  e(H)\equiv H (H^2)^{-1/2}= \gamma_5 D_W (D_W^{\dag} D_W)^{-1/2}
  \label{eq:4}
\end{equation}
where $H \equiv \gamma_5 D_W$ is a Hermitian operator.
Here we would like to focus directly on the operator $V$.  We will make
the assumption that $D_W^{\dag} D_W$ has no zero eigenvalues, so that
$(D_W^{\dag} D_W)^{-1/2}$ is well defined (it can be argued that such
zeroes are exceptional and irrelevant for the continuum limit).

Given any square matrix $M$ such that $M^{\dag} M$ has no zero eigenvalues, 
its ``polar decomposition'' is defined by
\begin{equation}
  M=U A
  \label{eq:5}
\end{equation}
where the matrix $U$ is unitary and the matrix $A$ is positive-definite
Hermitian~\cite{polar}.  It mirrors the expression of a complex number 
in terms of its phase and modulus.  The polar decomposition is unique,
with $A=(M^{\dag} M)^{1/2}$ and $U=M (M^{\dag} M)^{-1/2}$
 From Eq.~\ref{eq:3} we see that the matrix $V$ is the unitary factor
in the polar decomposition of the Wilson lattice operator $D_W$.
The main observation which we would like to make is that the unitary
factor $U$ in the polar decomposition of a matrix $M$ satisfies a 
maximum principle, namely $U$ is the unique matrix which maximizes
the expression ${\rm Re\; Tr}(U' M^{\dag})$, where $U'$ ranges
over the entire space of unitary matrices of the same size as $M$: 
\begin{equation}
  {\rm Re Tr} (U M^{\dag}) = {\rm Max}_{\{U' U'^{\dag} = I\}} 
  {\rm Re\; Tr}(U' M^{\dag})  
  \label{eq:6}
\end{equation}
The proof is straightforward.  We write $M$ in its polar form
and write $U'$ as $U'=U W$, where $W$ is also unitary.  This gives
\begin{eqnarray*}
  {\rm Re\; Tr}(U' M^{\dag})={\rm Re\; Tr}[U W (M^{\dag} M)^{1/2} U^{\dag}]
\end{eqnarray*}
\begin{equation}
  \qquad \qquad  ={\rm Re\; Tr}[W (M^{\dag} M)^{1/2}]    
  \label{eq:7}
\end{equation}
But $(M^{\dag} M)^{1/2}$ is a positive-definite Hermitian matrix
and we can therefore choose a basis where 
$(M^{\dag} M)^{1/2}={\rm Diag}(\lambda_1, \lambda_2, \dots \lambda_N),
\; \lambda_i>0$.
In this basis ${\rm Re\; Tr}[W (M^{\dag} M)^{1/2}] = \sum_{i}
{\rm Re} (w_{i,i}) \lambda_i$ and since the eigenvalues $\lambda_i$
are all positive and the unitarity of $W$ implies $|w_{i,i}| \le 1$,
the maximum occurs for $W=I$, i.e.~$U'=U$. Although this proof is very simple,
the property expressed by Eq.~\ref{eq:6} can hardly be found in books on
matrix algebra \cite{HH00}. Indeed very few pages, if any, are 
generally devoted to the polar decomposition itself.

Maximization and minimization principles have often proven very useful
in numerical analysis, since they can form the starting point for
efficient schemes of approximation.  Here we would like to briefly mention 
a few possibilities.  Assuming that one is interested in calculating
either $D \chi$ or $D^{-1} \chi$, one could construct the Krylov
space spanned by the vectors obtained acting with $H$ on $\chi$
(this space also underlies the approximation methods 
of~\cite{Neuberger2,EHN,Luescher2,Borici}):
\begin{equation}
  \chi_k = H^k \chi \qquad  k=0 \dots n
  \label{eq:8}
\end{equation}
In order to respect the symmetry properties of $V$ under $\gamma_5$
transformation, it is desirable to augment the basis by considering
separately the two projections
\begin{equation}
  \chi_{ks} = \frac{1 + s \gamma_5}{2} \chi_k  \qquad s=\pm 1
  \label{eq:9}
\end{equation}
For sufficiently large $n$ this basis would become overcomplete,
but we must work under the assumption that values of $n$ much
smaller than the dimensionality $N$ of the full vector space
will produce reasonable approximations.
We will finally denote by $\eta_{ks}$ an orthonormalized basis
in the space spanned by the vectors $\chi_{ks}$. (This will require
forming a number of scalar products of order $n^2$.  We are working
under the hypothesis that the values of $n$ one must consider make it
feasible both to calculate these scalar products and to store the
vectors $\eta_{ks}$.)  At this point the maximization of 
${\rm Re\; Tr}(U' M^{\dag})$ may be restricted to unitary operators
in the space $E$ spanned by the vectors $\eta_{ks}$, namely
\begin{equation}
  U'=\sum_{ks,k's'} v_{ks,k's'} \eta_{ks} \eta_{k's'}^{\dag} + I_{\bar E}
  \label{eq:10}
\end{equation}
where $I_{\bar E}$ denotes the identity in the complement of $E$.
Maximizing ${\rm Re\; Tr}(U' M^{\dag})$ one finds that the $(2n) \times (2n)$
matrix $v_{ks,k's'}$ is the unitary factor in the polar decomposition
of the truncated matrix $\eta_{k's'}^{\dag} D_W \eta_{ks}$.
As already mentioned above, this procedure will work only if reasonably
small values of $n$ can produce satisfactory approximations.

The truncation of Eq.~\ref{eq:10} does not represent, however, the
only way to take advantage of the maximization principle.  Another 
possibility is to assume an approximately ultralocal form for $V$. 
$V$ is known to be local \cite{Luescher2} and the approximation would 
consist in making it ultralocal.  Imposing such a condition on a 
unitary operator may be problematic, but one could write $V=\exp(\imath K)$
with $K^{\dag}=K$ and truncate to 0 the matrix elements of $K$
which exceed a maximum separation.  Otherwise, one could try to 
construct a better approximation to $V$ by the refinement of a 
first approximation: $V = V' \, V^{(0)}$.
A particularly appealing possibility is to take for $V^{(0)}$
the operator defined by a coarsening of the lattice obtained
by a blocking similar to those used in multigrid algorithms.
Multigrid techniques have been tried for lattice fermions,
but their application to the Wilson lattice operator~\cite{BERV}
turned out to be only marginally effective, most likely because of the 
presence of an additive mass renormalization.  Neuberger's operator is 
not affected by a similar renormalization and this offers the hope that 
multigrid methods may work much better.  In a multigrid approximation, 
one divides the lattice in cells of $2^d$ sites (for one level of
coarsening).  The gauge must be fixed within the cells, in order to
make the transport factors as close as possible to the identity.
Since this gauge fixing is local to the cells, it is not
computationally expensive.  The local gauge fixing leaves the freedom
of performing gauge transformations common to all the sites within the
cells.  This becomes a gauge symmetry of the blocked lattice.  After
the local gauge fixing it makes sense to define a projection over the
average cell fermionic fields.  We define new basis vectors
\begin{equation}
  \eta^{(0)}_X = 2^{-d/2} \sum_y \eta_{2X+y}
  \label{eq:11}
\end{equation}
where the vectors $\eta_x$ form a local basis for the fermions (spin 
and color indices are left implicit), $x$ and $X$ denote the integer
valued coordinates of the original and coarse lattices, respectively,
and $y$ are d-dimensional vectors with components that take value 0 or 1.
We can now define
\begin{equation}
  V^{(0)}=\sum_{XX'} v^{(0)}_{XX'} \eta^{(0)}_{X} \eta^{(0)\dag}_{X'}
  + I_{\bar E^{(0)}}
  \label{eq:12}
\end{equation}
where $I_{\bar E^{(0)}}$ denotes the identity in the complement of the
space spanned by $\eta^{(0)}_X$.  The matrix elements $v^{(0)}_{XX'}$
can then be determined by maximizing ${\rm Re Tr}(V^{(0)}
D_W^{\dag})$.  Alternatively, one could maximize ${\rm Re Tr}(V^{(0)}
D_W^{(0)\dag})$, where $D_W^{(0)}$ denotes a coarse lattice Wilson,
obtained by defining transport factors $U^{(0) \mu}_X$ between
neighboring cells as the $SU(3)$ matrices that maximize 
${\rm Re Tr}(U^{(0) \mu}_X S^{\mu\dag}_X)$, where $S^{\mu}_X$ denotes 
the sum of the transport factors $U^{\mu}_x$ connecting the boundary 
sites of the cells in the original lattice.  In both cases, the size 
of the matrix to determine is reduced by $2^d \times 2^d$.

Of course the crucial question is whether these or similar
computational procedures can provide a sufficiently accurate
approximation to Neuberger's operator without excessive usage of CP
time and memory.  We are currently working to test the approximations
outlined above and hope to report on our results in the near future.
At the same time, we hope that this communication may open new avenues
in the quest for efficient algorithms for lattice fermions.


\begin{thebibliography}{9}
\bibitem{Neuberger1}
H.~Neuberger, Phys.~Lett.~B417 (1998) 141.
\bibitem{GW}
P.~H.~Ginsparg, K.~G.~Wilson, Phys.~Rev.~D25 (1982) 2649.
\bibitem{Luescher1}
M.~Luescher, Phys.~Lett.~B428 (1998) 342.
\bibitem{HLN}
P.~Hasenfratz, V.~Laliena, F.~Niedermayer, Phys.~Lett.~B427 (1998) 125.
\bibitem{dw1}
D.~B.~Kaplan, Phys.~Lett.~B288 (1992) 342.
\bibitem{dw2}
 T.~Blum, A.~Soni, Phys.~Rev.~D56 (1997) 174.
\bibitem{dw3}
V.~Furman, Y.~Shamir, Nucl.~Phys.~B439 (1995) 54.
\bibitem{Neuberger2}
H.~Neuberger, Phys.~Rev.~Lett.~81 (1998) 4060.
\bibitem{EHN}
R.~G.~Edwards, U.~M.~Heller, R.~Narayanan, Nucl.~Phys.~B540 (1999) 457
\bibitem{Luescher2}
P.~Hernandez, K.~Jansen, M.~Luescher,
Nucl.~Phys.~B552 (1999) 363.
\bibitem{Borici}
A.~Bori\c ci Phys.~Lett.~B 453 (1999) 46.
\bibitem{polar}
see e.g. in R.~Bathia, ``Matrix Analysis'', Graduate Texts in
Mathematics Vol.~169, Springer-Verlag, Berlin, 1996
\bibitem{HH00}
Eq.~\ref{eq:6} is mentioned in
``Numerische Methoden der linearen Algebra'' by K.~D.~Faddejew and
W.~N.~Faddejewa (VEB, Berlin 1976) and 
Ky~Fan and A.~J.~Hoffmann, Proc.~Amer.~Math.~Soc. 6 (1955) 111.
We wish to thank R.~Horsley and H.~Neuberger for bringing these references
to our attention.
\bibitem{BERV}
R.~Brower, R.~Edwards, C.~Rebbi and E.~Vicari
Nucl.~Phys.~B336, (1991) 689.
\end{thebibliography}
\end{document}